\documentclass[12 pt]{article}
\usepackage{rotating}
\usepackage{graphicx, caption, subfig,amsmath, enumitem, setspace,graphics, comment, amstext,lscape, hyperref}
\usepackage[usenames, dvipsnames]{color}
\usepackage[longnamesfirst]{natbib}
\usepackage{rotating}
\bibpunct{(}{)}{;}{author-year}{}{,}
\usepackage[margin=9pt,font=small,labelfont=bf,labelsep=endash]{caption}
\usepackage{booktabs}

\title{Photographic home styles in Congress: a computer vision approach} 

\author{L. Jason Anastasopoulos\thanks{(First Author, Corresponding Author); Assistant Professor, Department of Political Science and Department of Public Administration and Policy, School of Public and International Affairs, University of Georgia;  Address: 355 S. Jackson St., Athens, GA 30602, USA;  Email: \href{ljanastas@uga.edu}{ljanastas@uga.edu}.; Tel: +001 973.641.8258}.
 \and
	Dhruvil Badani\thanks{BS Undergraduate Student, Department of Computer Science, UC Berkeley; Address: 1263 Trestlewood Lane, San Jose, CA 95138, USA; Email: \href{dhruvilb@berkeley.edu}{dhruvilb@berkeley.edu}.}
	\and 
	Crystal Lee\thanks{Software Engineer at Pinterest.com; BS in Computer Science, UC Berkeley (2015);  Address: 455 Cloverdale Ct, Sunnyvale CA 94087, USA;  Email: \href{isealya@gmail.com}{isealya@gmail.com}.}
	\and
	Shiry Ginosar\thanks{Doctoral student, Department of Computer Science, UC Berkeley; Address: UCB EECS, MC 1776
387 Soda Hall, Berkeley, CA 94720-1776, USA;  Email: \href{shiry@eecs.berkeley.edu}{shiry@eecs.berkeley.edu}}.
		\and
	Jake Ryland Williams\thanks{Assistant Professor, Department of Information Science, Drexel University; Address: 30 N. 33rd Street
Philadelphia, PA 19104;  Email: \href{jakerylandwilliams@gmail.com}{jakerylandwilliams@gmail.com}.}
	}

\begin{document}
\maketitle

\begin{abstract}
While members of Congress now routinely communicate with constituents using images on a variety of internet platforms, little is known about how images are used as a means of strategic political communication.  
This is due primarily to computational limitations which have prevented large-scale, systematic analyses of image features.
New developments in computer vision, however, are bringing the systematic study of images within reach.
Here, we develop a framework for understanding visual political communication by extending Fenno’s analysis of home style ~\citep{fennohomestyle} to images and introduce ``photographic'' home styles.
Using approximately 192,000 photographs collected from MCs Facebook profiles, we build machine learning software with convolutional neural networks\footnote{\ This software was originally created in \textbf{Python} using the deep learning framework \textbf{Caffe}, we plan on releasing an \textbf{R} version of the software developed for this paper upon publication.} and conduct an image manipulation experiment to explore how the race of people that MCs pose with shape photographic home styles.
We find evidence that electoral pressures shape photographic home styles and demonstrate that Democratic and Republican members of Congress use images in very different ways.

\end{abstract}

\doublespacing
\section*{Introduction}

Before becoming president in 1963 after John F. Kennedy's assassination, Lyndon Johnson spent 12 years as a Democratic representative of Texas' 10th Congressional district and another 12 years representing Texas as a Senator.
During his time in Congress, he voted against practically every piece of pro-civil rights legislation~\citep{caro1990means,caro2002master} yet became known as one of America's foremost civil rights champions shortly after signing the landmark Civil Rights Act of 1964 into law~\citep{kotz2005judgment}.

\begin{figure}[!h]
	\centering
	\includegraphics[width = .75\textwidth]{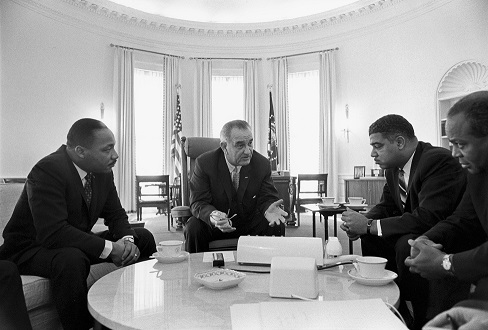}
	\caption{Johnson meets with Martin Luther King and other civil rights leaders in the oval office. \textit{Source}: Lyndon Baines Johnson Library and Museum.  \textit{Photo by Yoichi Okamoto}.}
	\label{fig:mlk}
\end{figure}

\begin{figure}[!h]
	\centering
	\includegraphics[width = .75\textwidth]{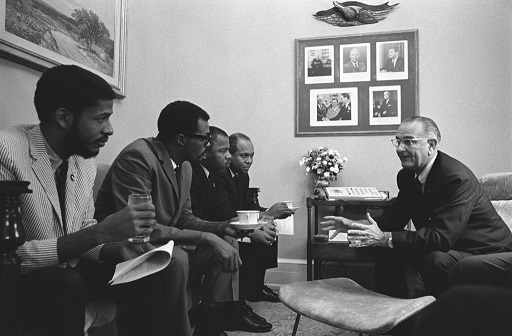}
	\caption{President Johnson meets with civil rights activists John Lewis and James Farmer.\textit{Source:} \href{http://www.pbs.org/newshour/rundown/obama-pays-tribute-civil-rights-act-50th-anniversary/}{http://www.pbs.org/newshour/rundown/obama-pays-tribute-civil-rights-act-50th-anniversary/}. \textit{Photo by Yoichi Okamoto}.}
	\label{fig:civrights}
\end{figure}

The transformation from perceived civil rights opponent to civil rights leader in such a short period of was time due in no small part to Johnson's skillful abilities to manipulate how the public and those with influence perceived him through photography~\citep{altschuler1986lyndon,zarefsky2004presidential,duganne2013photographic}. 
For example, Johnson made it a point to invite photographers into the oval office and elsewhere to take pictures of meetings with African-American civil rights leaders and activists such as Martin Luther King (Figure~\ref{fig:mlk}) and others (Figure~\ref{fig:civrights}).
Indeed, LBJ was the first president to appoint an official White House photographer, Yoichi Okamoto, now known as the ``godfather'' of White House photography, who served under him from 1963 to 1969, snapping now iconic shots of Johnson conducting business in the White House~\citep{estrinphoto}. 

Okamoto's photos are considered masterful not only because of their aesthetic qualities such as lighting, angles, etc., but also because they are believed to have perfected a visual persuasion technique known as ``juxtaposition''~\citep{benjamin1972short,sontag1977photography,sigurjonsdottir2016hot} which simply involves placing an individual or object next to another object, person or group in a photograph.
In political contexts, juxtaposition is used as a means of creating associations between political figures with something that an object, person or group represents.
In Figures~\ref{fig:mlk} and~\ref{fig:civrights}, for example, the goal of juxtaposing Johnson with civil rights leaders was to associate him with the civil rights movement.
While photographs of political figures are replete with intentional and unintentional juxtapositions, there is little empirical evidence demonstrating how or if this technique and others used in photography affect how political figures are perceived by the public.

Developing an empirical basis for measuring the political uses and effects of visual persuasion techniques such as juxtaposition is crucial to understanding key aspects of the photo-laden modern day political landscape.
With the explosion of social media as a means of constituency communication and improvements in camera technology, text posts combined with photographs and images have become a routine means of daily communication between members of Congress, their constituents and the public at large through a number of online platforms such as Twitter, Facebook and Instagram and MCs own \textit{.gov} web pages hosted on U.S. government servers~\citep{adler1998home,esterling2012connecting,grimmer2013representational,gulati2013social,peterson2012tweet}.
In the United States, juxtapositions of political figures with members of constituency groups are particularly relevant to understanding  present day home styles of members of the U.S. House of Representatives and Senate~\citep{fennohomestyle} because they can help us understand how members of Congress strategically utilize image-rich new media to convey key elements of MCs home styles.

This study has three goals.  
First, we develop a framework for understanding strategic political communication through visual media such as photos and images by extending Fenno's analysis of home style ~\citep{fennohomestyle} and introduce the concept of a ``photographic'' home style. 
Next, we demonstrate that image features identified in our framework causally affect how politicians are perceived by the public on dimensions related to home style using an image manipulation experiment.
Finally, we study the photographic home styles~\citep{fennohomestyle} of House and Senate members by comparing racial demographics of the photographs that they post on Facebook with the racial demographics of their respective constituency groups.
To accomplish this, we collected 192,000 photographs from the Facebook profiles of 230 members of the House of Representatives and 52 members of the Senate and build software based on a machine learning technique known as convolutional neural networks (CNNs) to identify the race of each individual in members' Facebook profile photos.

The remainder of this paper is organized as follows. 
Section~\ref{framework} describes a general framework for analyzing political images; Section~\ref{photostyles} introduces the concept of ``photographic'' home styles by applying the framework discussed in section~\ref{framework}; Section~\ref{experiment} presents an experiment which demonstrates how features of photographs causally affect perceptions of politicians related to elements of photographic home styles discussed in section~\ref{photostyles}; Section~\ref{cnns} discusses an empirical analysis of photographic home styles using a convolutional neural network and Section~\ref{discussion} concludes with a discussion of the implications of our findings and possible directions for future research in this area.

\section{A Framework for Political Image Analysis} \label{framework}

A brief exploration of the literature on the political uses of images suggests that what distinguishes a political image from a non-political image is that political images are usually created with the intention of persuading viewers to side with a political candidate, party or cause~\citep{barrett2005bias, rosenberg1991creating}.  

\begin{figure}[!h]
	\centering
	\includegraphics[width =\textwidth]{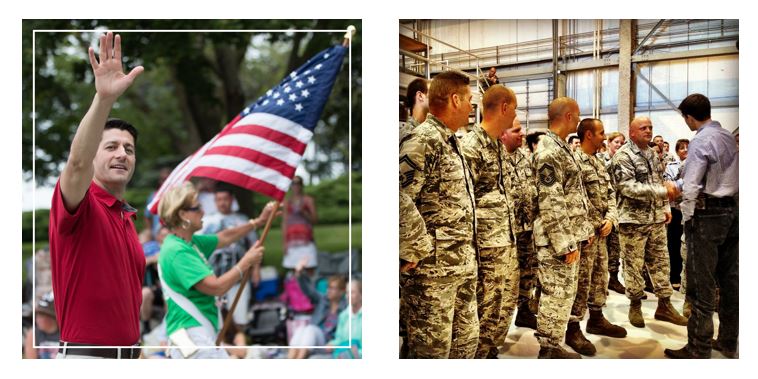}
	\caption{Photos from Speaker Paul Ryan's Facebook profile showing him with a flag in the background (left) and speaking to military personnel (right). \textit{Source:} \href{https://www.facebook.com/paulryanwi/}{https://www.facebook.com/paulryanwi/}}
	\label{fig:ryanpics}
\end{figure}

\begin{figure}[!h]
	\centering
	\includegraphics[width =\textwidth]{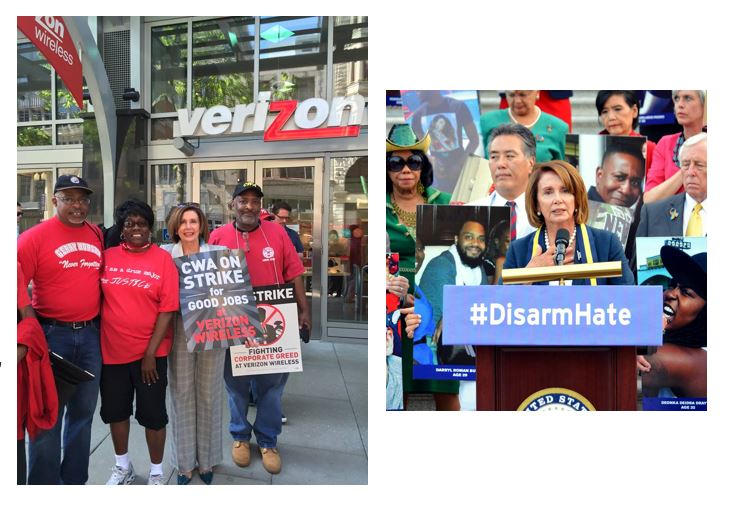}
	\caption{Pictures from minority leader Nancy Pelosi's Facebook profile showing her posing next to union members on strike in her district (left) and speaking at a gun control rally (right). \textit{Source:}\href{https://www.facebook.com/NancyPelosi/}{https://www.facebook.com/NancyPelosi/}}
	\label{fig:pelosipics}
\end{figure}

Starting with politicians' use of images, the focus of our empirical study below,  we premise our framework on the Mayhewian electoral connection assumption that the behavior of democratically elected politicians can be largely explained by their desire to get re-elected~\citep{mayhew1974congress}.
Under this premise, images are used by politicians to increase their likelihood of re-election or otherwise advance their careers.
The primary goal of political image analysis, then, is to understand \textit{how} politicians use photographs to fulfill their re-election goals or advance their careers. 

On this front, Fenno's analysis of legislator ``home style''~\citep{fennohomestyle} provides guidance. 
According to Fenno, home style is an idiosyncratic set of behaviors legislators adopt as a means of fostering trust among their constituents for the purpose of increasing their chances re-election.
Legislator home style is comprised of three main elements: (1) how she allocates resources; (2) how she presents herself to others (presentation of self) and (3) how she explains what she is doing when outside of her district (Washington activities). 
As we describe in more detail below, photographs provide an excellent means of communicating presentation of self and Washington activities to constituents.

For Republican House members, this might involve posting photographs on social media posing with American flags in the background or addressing military personnel (Figure~\ref{fig:ryanpics}). 
For Democratic House members, this might involve posing with union members in their districts or giving speeches on gun control as in Figure~\ref{fig:pelosipics}.

With this in mind, we first identify and discuss three essential features of photos which can shape how politicians are perceived: objects, people and poses.

\begin{figure}[!ht]
	\centering
	\includegraphics[width =.8\textwidth]{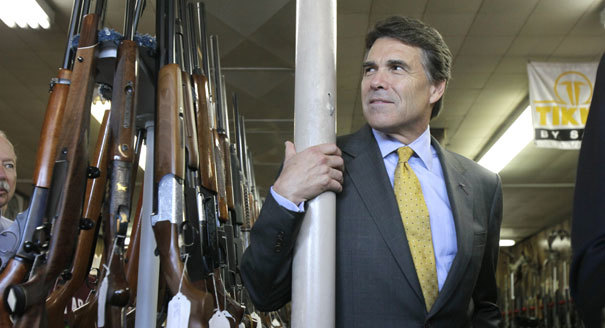}
	\caption{Former Texas Governor Rick Perry posing with rifles in a gun shop. \textit{Source} \href{http://www.politico.com/gallery/politicians-with-guns}{http://www.politico.com/gallery/politicians-with-guns}}
	\label{fig:perrygun}
\end{figure}

\subsection{Objects}
Objects and symbols in photos contain political meaning when juxtaposed with a political figure or when they are displayed by themselves.
Objects tend to contain abstract information about the personal qualities of politicians and/or their policy positions. 
For example, the Democratic donkey and Republican elephant are recognizable by many as representing American political parties but also stand for the \textit{ideas} and \textit{policy positions} taken by each party. 
The American flag, an object and a symbol, represents the geopolitical construct of the United States but depending on context can represent American values and ideals, American patriotism, American military power or a combination of all three.

Objects such as guns and military equipment can be used to represent specific policy stances such as support or opposition to gun control, support for veterans, military interventions and so on.
Thus, a politician might convey opposition to gun control and/or support for the Second Amendment by posing in a gun shop as former Republican Texas Governor Rick Perry does in Figure~\ref{fig:perrygun}.

\subsection{People}

\begin{figure}[!h]
	\centering
	\includegraphics[width =\textwidth]{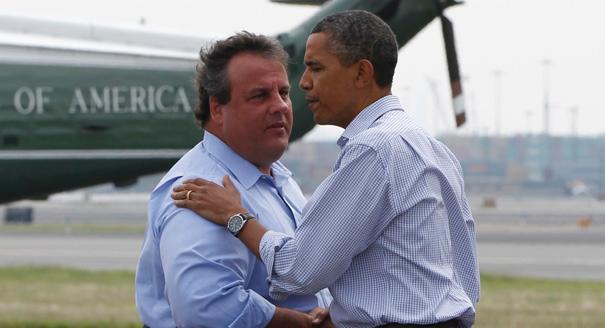}
	\caption{New Jersey governor Chris Christie embraces President Obama during his visit to the state after Hurricane Sandy in 2013. \textit{Source} \href{http://www.breitbart.com/big-government/2015/07/01/inside-hug-gate-the-online-meme-that-chris-christie-cant-shake/}{http://www.breitbart.com/big-government/2015/07/01/inside-hug-gate-the-online-meme-that-chris-christie-cant-shake/}}
	\label{fig:christieobama}
\end{figure}

Other people in photographs where a politician is the subject are politically relevant to the extent that they can convey information about the politician to their constituents. 
This can be accomplished through posing with a well-known person (celebrity, political figure, etc.) or unknown people. When posing with unknown individuals, visible features (race, gender, veteran status, age etc.) of these individuals can shape how the politician is perceived.
For example, a white politician who consistently poses with people from other racial groups may be trying to signal that they identify or empathize with members of that group or that their policy positions will benefit members of that group.
In our image experiment below, we demonstrate that the race of individuals that politicians pose with affects how they are perceived by politicians in terms of partisanship/party identification and identification and empathy with minority groups.

Regarding famous individuals, the information conveyed depends largely upon how the famous individual is perceived by the viewer. For a recent example, Figure~\ref{fig:christieobama} contains a photo of Republican New Jersey governor Chris Christie embracing President Obama after a visit to assess some of the damage caused by Hurricane Sandy in 2013. 
While right-leaning media sources such as Breitbart excoriated Christie, mainstream sources praised Christie's bi-partisanship\footnote{
The \textit{New York Times}, for example, covered the meeting between Obama and Christie with a story whose headline read ``No Partisan Fire at the Shore: An Obama-Christie Reunion.''\textit{Source: The New York Times, \href{http://www.nytimes.com/2013/05/29/us/obama-and-christie-to-view-recovery-on-jersey-shore.html?\_r=0}{http://www.nytimes.com/2013/05/29/us/obama-and-christie-to-view-recovery-on-jersey-shore.html?\_r=0}}}.

\subsection{Poses}

Poses, which include actions and facial expressions, are among the most well studied aspects of visual persuasion. 
In a recent study by~\citep{joo2014visual}, for example, the authors find that certain facial expressions and actions convey both positive and negative politically relevant information on a number of dimensions.
Facial displays such as smiles and frowns can convey satisfaction or sadness with a certain political event or outcome.
Gestures such as finger pointing, hand waving, hand shaking, hugging and so on can demonstrate different types of engagement with constituents and other political leaders.

\section{Photographic Home Styles} \label{photostyles}

As we mentioned above, home styles are the idiosyncratic sets of behaviors that members of Congress adopt to gain trust among their constituents and are comprised of two components that can be represented visually: Washington activities and presentation of self. 
``Photographic'' home styles, then, refer to how the politically relevant features of photographs discussed above (objects, people and poses), visually reflect Washington activities and presentation of self.

\begin{figure}[!h]
	\centering
	\includegraphics[width = \textwidth]{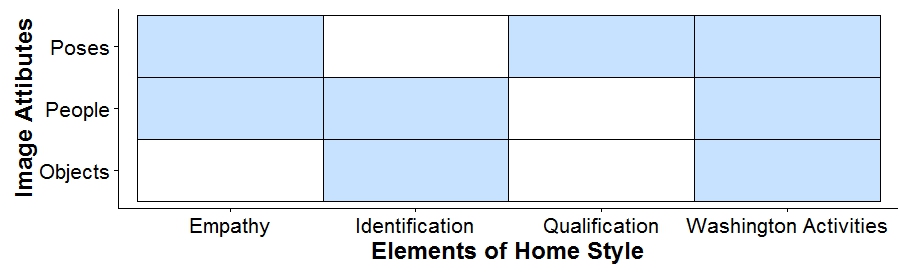}
	\caption{Image attributes which reflect elements of photographic home style. \textit{Blue $=$ Reflects home style element; White $=$ Does not reflect home style element.}}
	\label{fig:imageactions}
\end{figure}

Figure~\ref{fig:imageactions} breaks down each aspect of home style and identifies which features of images correspond to each aspect.
Our theory of photographic home style suggests that each image feature reflects clusters of home style aspects. 
Objects, for example, can represent identification with constituents and Washington activities, but are unlikely to represent empathy and qualification.

Other people in photographs, the focus of our experiment and empirical study below, can convey empathy or identification with constituent groups.
For example, photographs in which members of Congress pose with veterans or members of minority groups communicate that the MC cares about members of that group, empathizes with them or at the very least spends time with them. 
We provide concrete evidence for this in our image experiment described below and discuss how image features relate to each aspect of home style in more detail.  

As mentioned above, explanations of MCs activities in Washington increase their re-election chances though fostering trust among constituents (Fenno 1978).
One method of gaining trust through Washington activities involves demonstrations by MCs that they are ``different'' from or better than other members of Congress,~\citep{fennohomestyle,fiorina1991home}. 
Fenno notes that ``members of Congress run for Congress by running against Congress. The strategy is ubiquitous, addictive, cost free and foolproof''~\citep{fennohomestyle}.

\begin{figure}[!h]
	\centering
	\includegraphics[width = .5\textwidth]{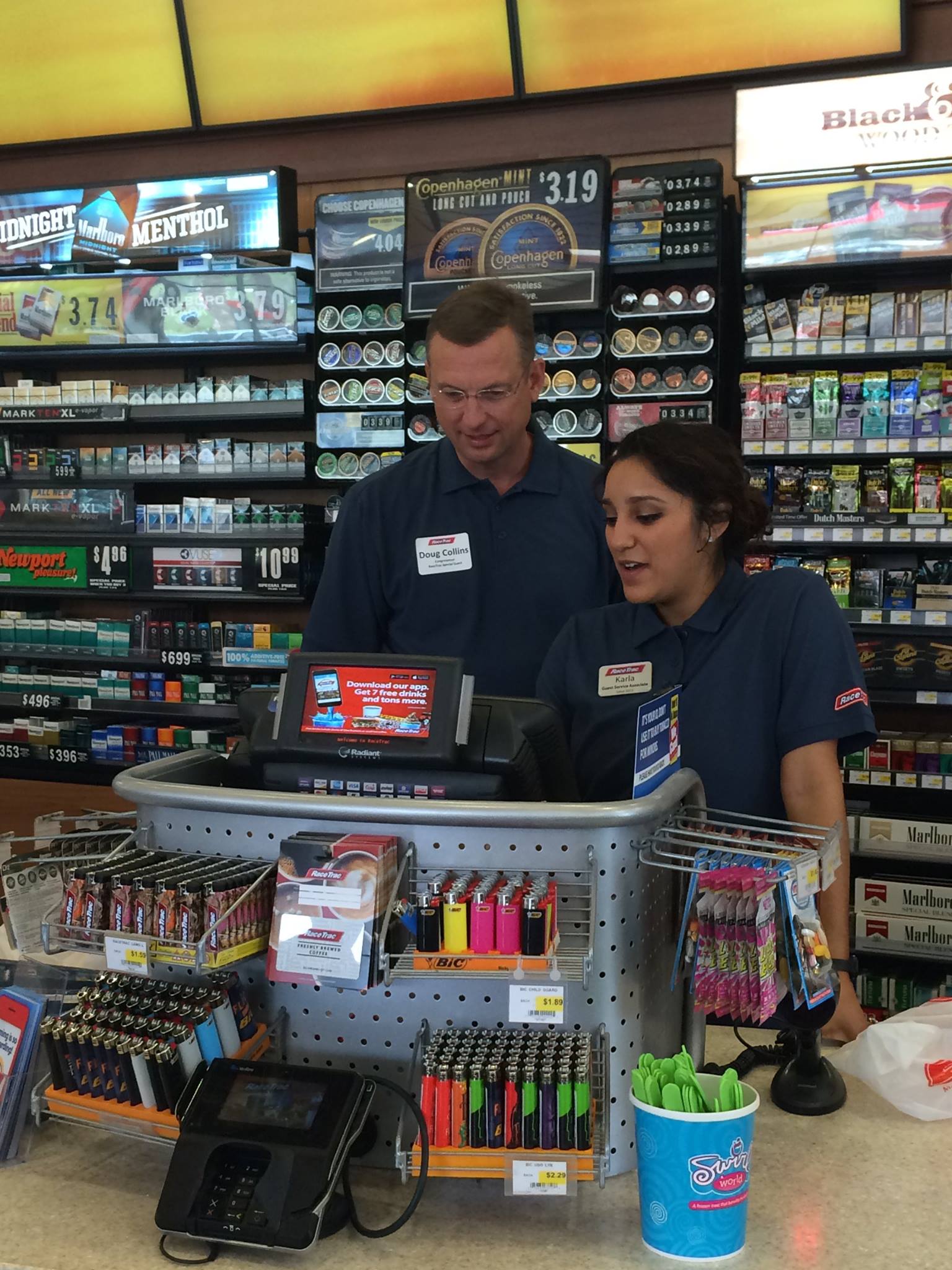}
	\caption{Rep. Doug Collins (R-GA 9, right) posing as a convenience store clerk in his district. \textit{Source}: Rep. Collins' Facebook page: \href{https://www.facebook.com/RepresentativeDougCollins/photos/}{https://www.facebook.com/RepresentativeDougCollins/photos/}}
	\label{fig:dougcollins}
\end{figure}

This strategy is readily apparent in photos. 
For example, many representatives contrast themselves with the ``Washington elite'' by posing as blue-collar workers as Rep. Doug Collins (R-GA 9) does in Figure~\ref{fig:dougcollins} where he poses as a clerk at a gas station in his district.

\begin{figure}[!h]
	\centering
	\includegraphics[width = .75\textwidth]{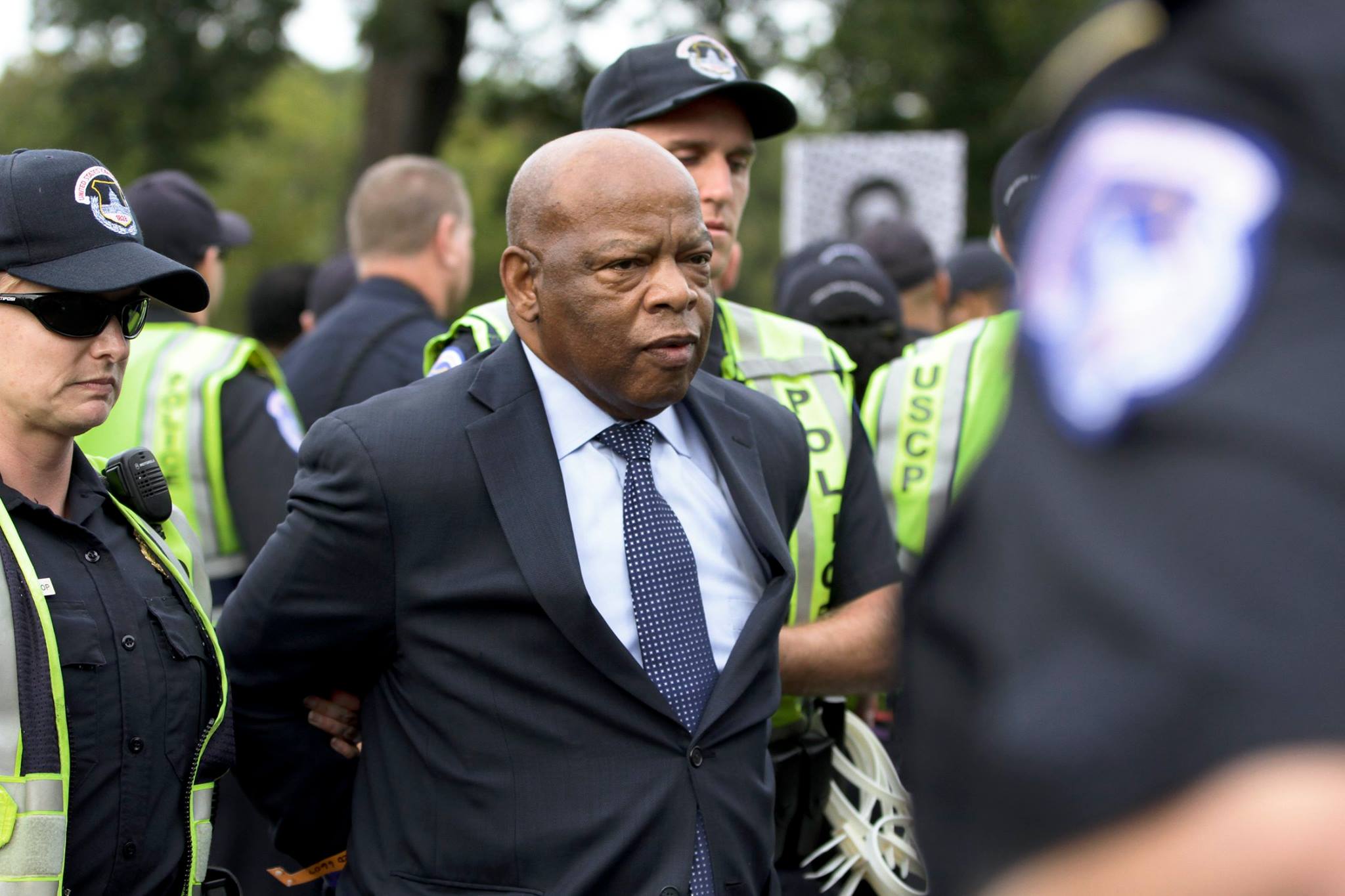}
	\caption{Rep. John Lewis (D-GA 5, center) being arrested for blocking traffic during an immigration rally in Washington DC in 2012. \textit{Source}: Rep. Lewis' Facebook page: \href{https://www.facebook.com/RepJohnLewis/photos}{https://www.facebook.com/RepJohnLewis/photos}}
	\label{img:johnlewis}
\end{figure}

More conspicuous means of explaining Washington activities through photos include shots in which MCs are clearly \textit{in} Washington DC speaking to large crowds, meeting with other political leaders, voting on bills or advocating for policy change. 
A more extreme example of this is Figure~\ref{img:johnlewis} which contains a pose of Rep. John Lewis (D-GA 5) being arrested by DC police for blocking traffic during an immigration reform rally in 2012.

\begin{figure}[!h]
	\centering
	\includegraphics[width = .5\textwidth]{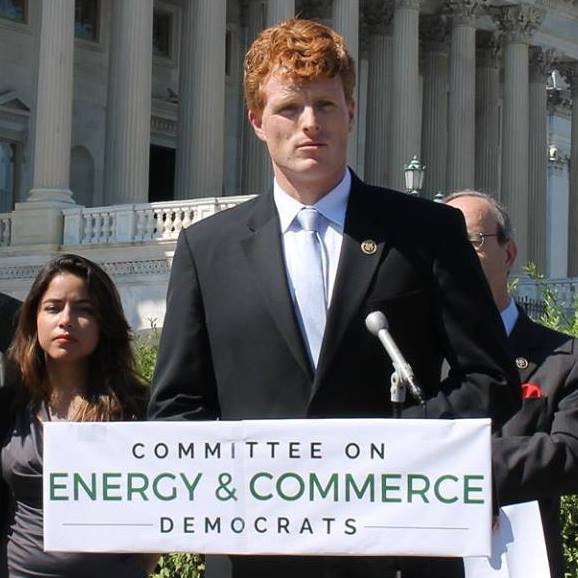}
	\caption{Rep. Joe Kennedy III (D-MA 4) giving a speech on behalf of the Energy and Commerce Committee. \textit{Source}: Joe Kennedy III Facebook page: \href{https://www.facebook.com/CongressmanJoeKennedyIII/}{https://www.facebook.com/CongressmanJoeKennedyIII/}}
	\label{img:joekennedy}
\end{figure}

Presentations of self, which include demonstrations of competence (qualification), identification and empathy with constituents are also readily conveyed through objects, people and poses in photographs. 
Competence, for example, is demonstrated visually in several ways. 
Photos communicating leadership ability can might include any situation in which MCs are speaking to large crowds or at a podium containing some markers of prestige. 
An example of this is Rep. Joe Kennedy III's (D-MA 4) Facebook profile photo (Figure~\ref{img:joekennedy}) taken while he was giving a speech on behalf of the Energy and Commerce Committee. 

Demonstrations of identification and empathy through photographs convey the message to constituents that I am ``one of you'' and ``I understand you and think like you do.'' 
Photos which show MCs participating in activities that their constituents enjoy such as hunting, attending sporting events, performing manual labor etc are one means of accomplishing this. 

\begin{figure}[!h]
	\centering
	\includegraphics[width = .6\textwidth]{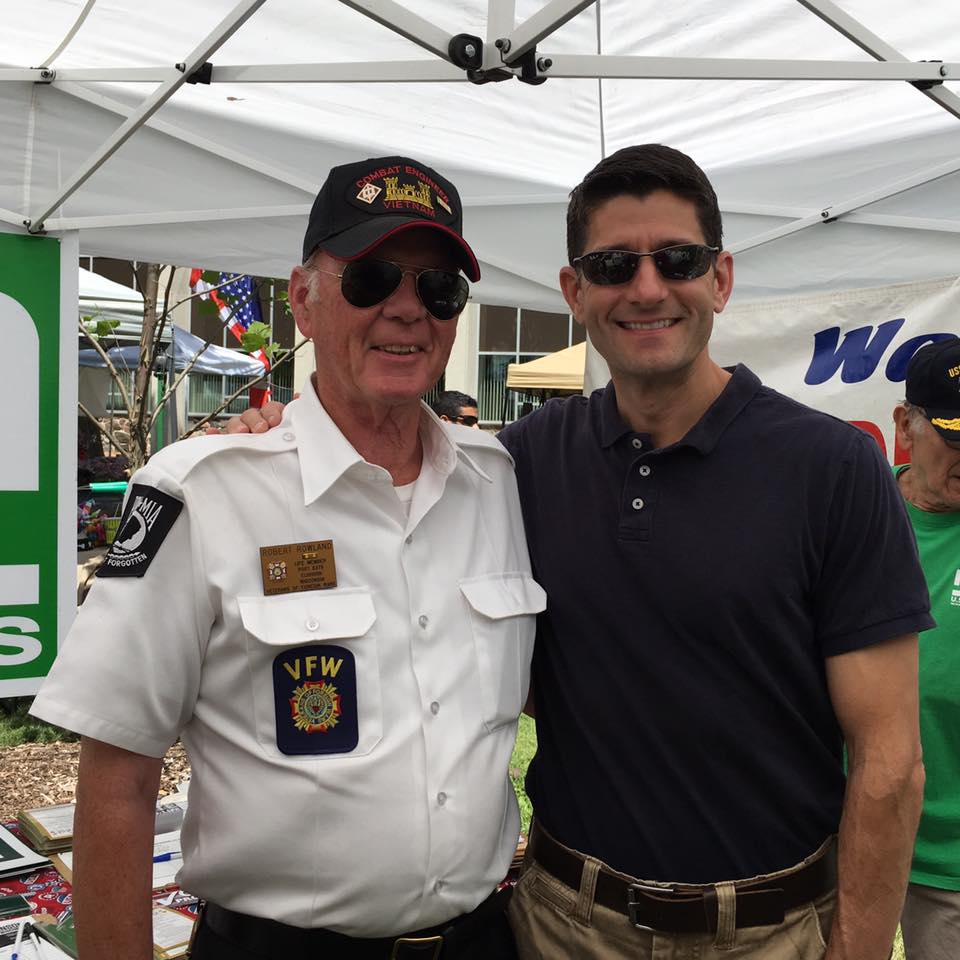}
	\caption{House Speaker Paul Ryan Poses with a Vietnam Veteran. \textit{Source:} \href{https://www.facebook.com/paulryanwi/}{https://www.facebook.com/paulryanwi/}.}
	\label{img:ryanveteran}
\end{figure}

Another means is through posing with members of constituency groups that are likely to vote for the MC in the next election.
If this were true, we would expect Republicans to strategically post pictures with members of groups that tend to vote Republican, such as veterans as we see with House Speaker Paul Ryan in Figure~\ref{img:ryanveteran}.

\begin{figure}[!h]
	\centering
	\includegraphics[width = .75\textwidth]{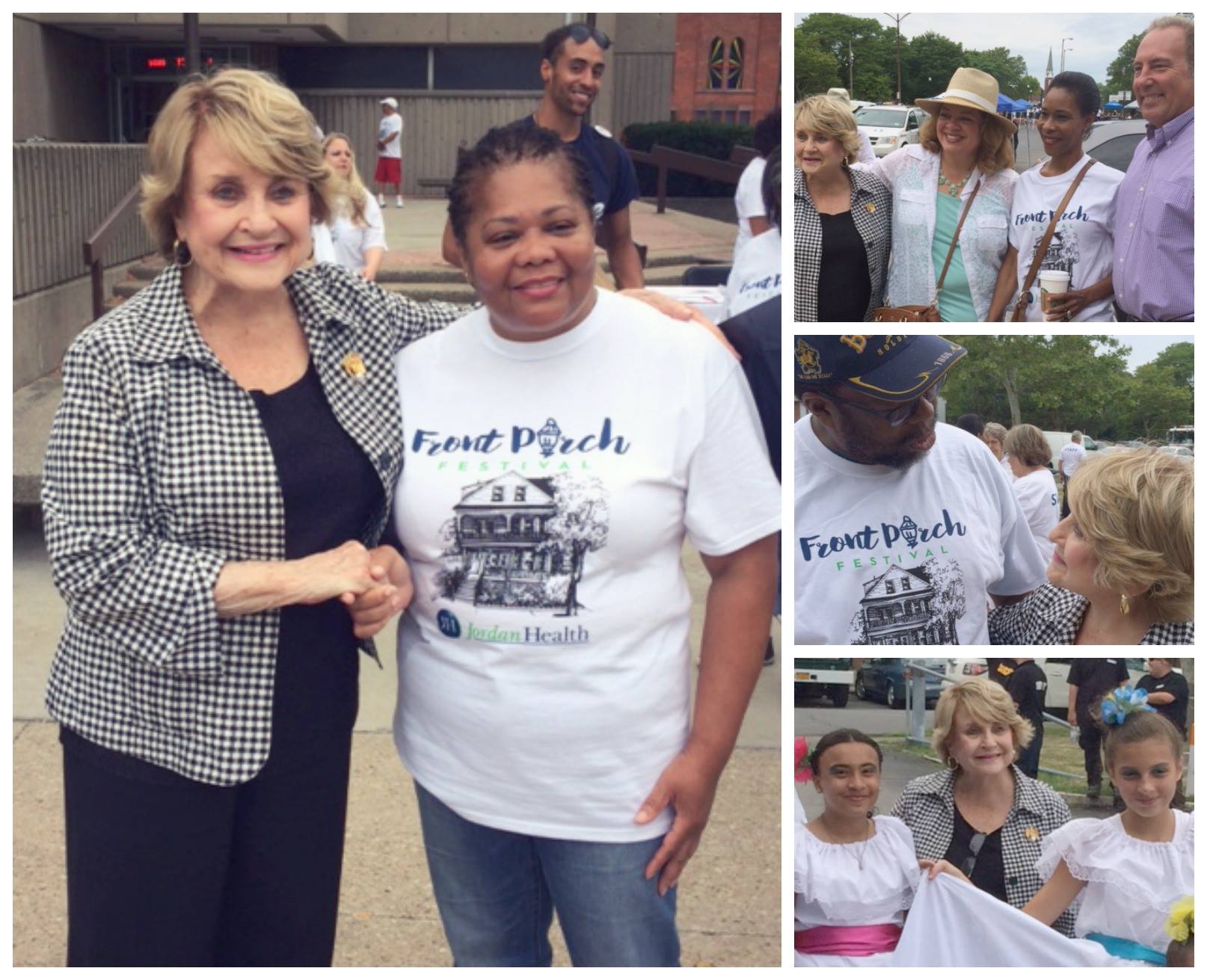}
	\caption{A photo collage posted on Rep. Louise Slaughter's (D-NY 25) Facebook profile where she is posing with African-American, Hispanic and white constituents at a festival in her district. \textit{Source:}\href{https://www.facebook.com/RepLouiseSlaughter/}{https://www.facebook.com/RepLouiseSlaughter/}.}
	\label{img:repslaughter}
\end{figure}

Similarly, Democrats would be more likely to do so by posting pictures of themselves with members of minority groups such as African-Americans and Hispanics, members of labor unions and so forth.
Rep. Louise Slaughter (D-NY 25) provides an good example of this in a photo collage posted on her Facebook profile page in Figure~\ref{img:repslaughter} where she meets with with African-American, Hispanic and white members of her district. 

As we demonstrate with an empirical study below, we find strong evidence of an electoral connection with photographic home styles.
While Republican MCs appear to make little effort to represent Hispanic and African-Americans in their profile photos, we find evidence that Democratic members of the House and Senate strategically match the minority proportions in their districts with the minority proportions in photos. 

\cleardoublepage

\section{The people you pose with experiment}\label{experiment}

\begin{figure}[!ht]
	\centering
	\includegraphics[width = \textwidth]{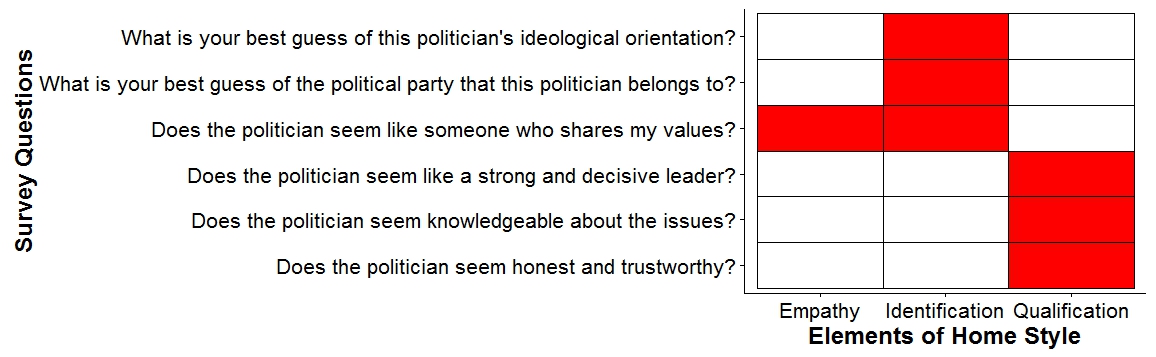}
	\caption{Questions asked to respondents and their relationship to each element of home style. }
	\label{fig:exptoutcomes}
\end{figure}

The empirical portion of our paper below explores the connection between racial representation and photographic home styles through a systematic examination of the race of individuals that members of Congress pose with and the photographs they choose to post on their social media sites.
Before moving on to this analysis, however, we first seek to demonstrate that group characteristics such as the race and gender of people that members of Congress pose with causally affect politically relevant outcomes related to home style. 
To accomplish this, we designed an image manipulation experiment in which the treatment randomized was a series of images containing a relatively obscure member of Congress posing by himself (control) or next to people of different genders and races which was presented to survey respondents\footnote{ See Appendix C for details about subject recruitment and survey design.}.

Outcomes collected were respondents ``best guess'' about the MCs qualities that directly relate to presentations of self relevant to  qualification, identification and empathy as shown in Figure~\ref{fig:exptoutcomes}.
No information other than the photograph and a sentence explaining that the person in the photograph is a politician in the United States was provided to respondents. 

\begin{figure}[!ht]
	\centering
	\includegraphics[width=\textwidth]{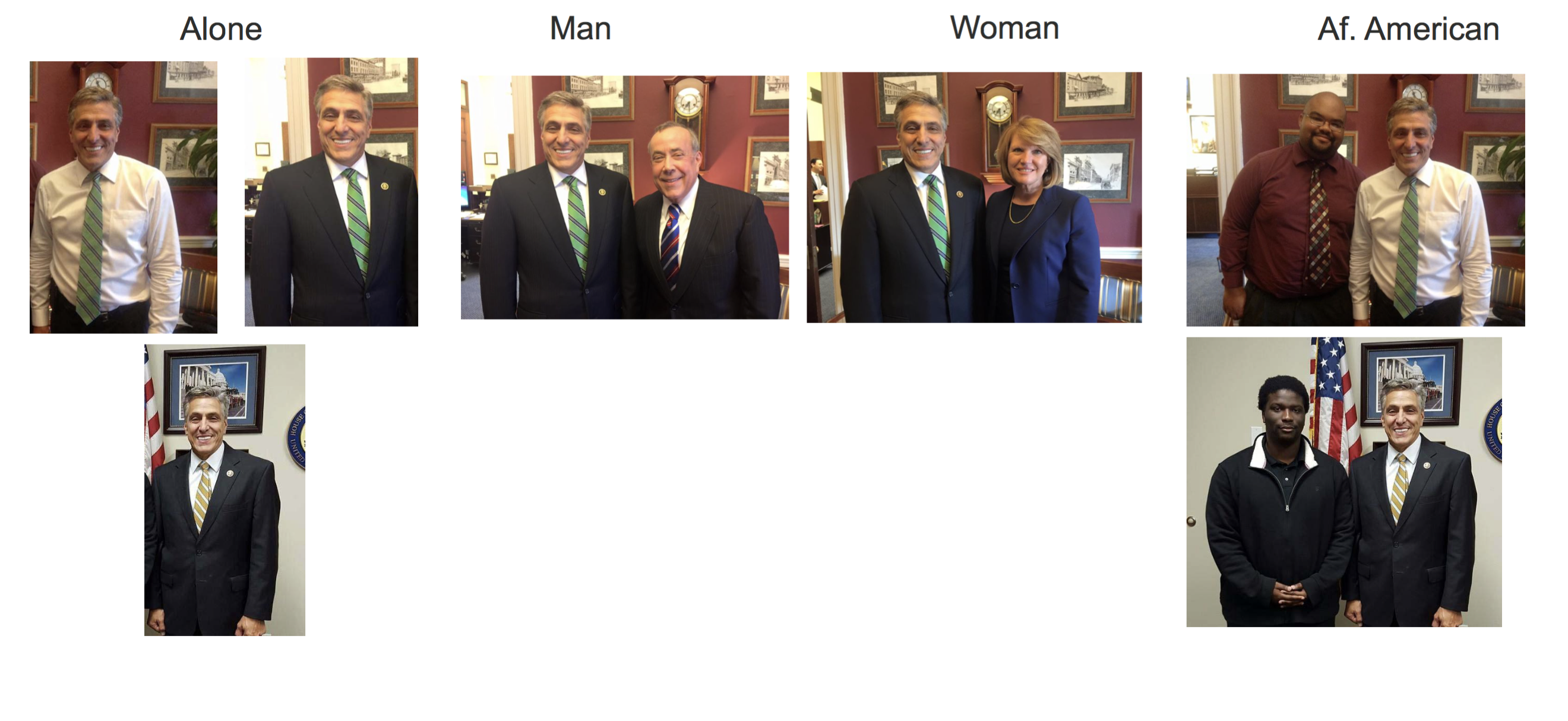}
	\caption{Experimental treatments shown to subjects taken from Facebook photos.}
	\label{fig:treats}
\end{figure}

Figure~\ref{fig:treats} contains each of the potential image treatments that respondents were randomly exposed to.  
The political subject of the experiment that was chosen was  Republican Rep. Lou Barletta who represents Pennsylvania's 11th District. 
Rep. Barletta was chosen for several reasons. 
First, he is a relatively obscure political figure\footnote{ Less than 1\% of survey respondents recognized Barletta.}, thus making it unlikely that respondents will base their judgments about him based on anything outside of the images.
Second, he happens to have many pictures on his Facebook profile where he is posing with people of different races and genders with the same background, allowing us to isolate the effects of the race and gender of the person standing next to him on the outcomes shown in Figure~\ref{fig:exptoutcomes}.

\begin{figure}[!ht]
	\centering
	\includegraphics[width=.7\textwidth]{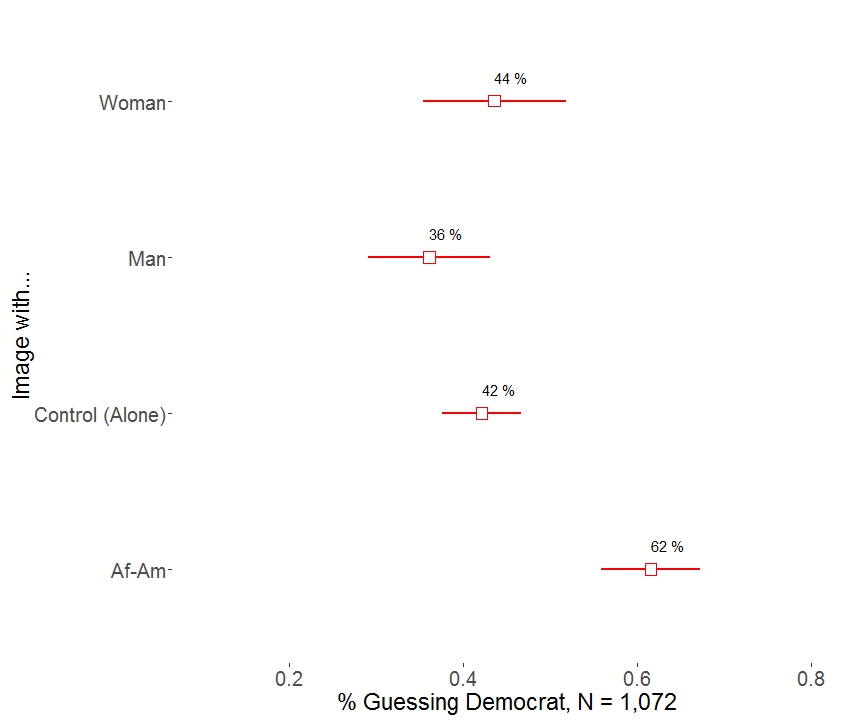}
	\caption{``What is your best guess of the political party that this politician belongs to?'' \% of respondents guessing is a Democrat with 95\% confidence intervals.}
	\label{fig:demguess}
\end{figure}

\begin{figure}[!ht]
	\centering
	\includegraphics[width=.7\textwidth]{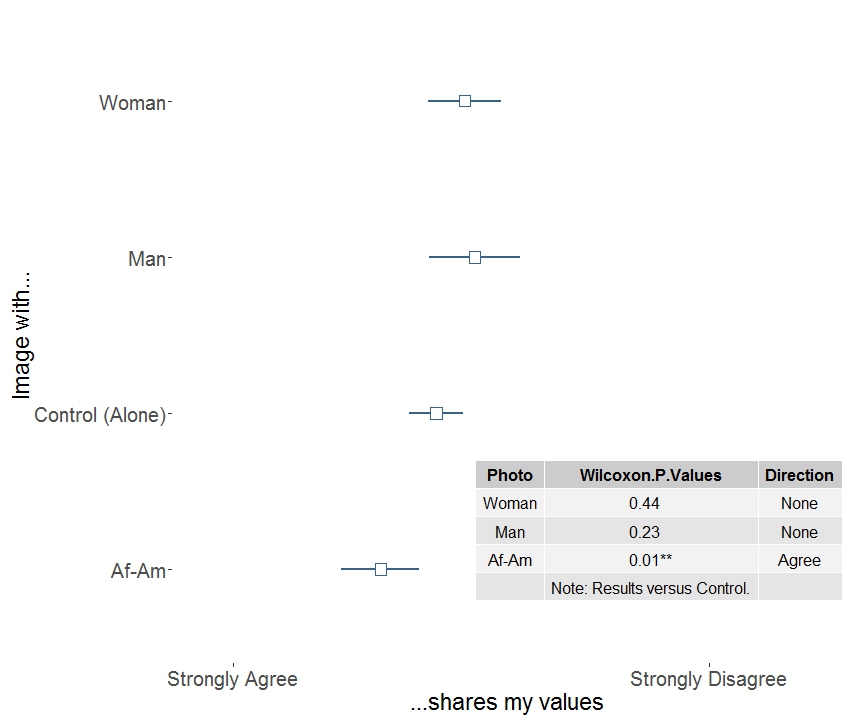}
	\caption{``Does the politician seem like someone who shares my values?'' Average agreement among non-white respondents with 95\% confidence intervals. Wilcoxon p-values under $H_{0}: \mu_{Control} - \mu_{Treatment} = 0$}.
	\label{fig:sharevalues}
\end{figure}

\begin{figure}[!ht]
	\centering
	\includegraphics[width=.7\textwidth]{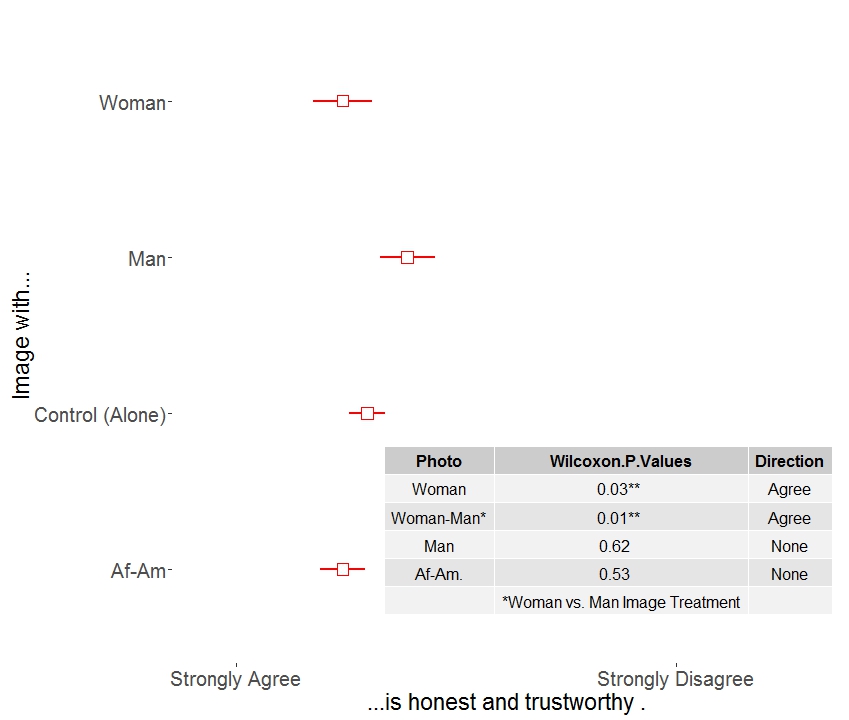}
	\caption{``Does the politician seem honest and trustworthy?'' Average agreement among respondents with 95\% confidence intervals. Wilcoxon p-values under $H_{0}: \mu_{Control} - \mu_{Treatment} = 0$}
	\label{fig:trustworthy}
\end{figure}

We find treatment effects for race and gender across different categories of political outcomes. 
Posing next to African-Americans made respondents more likely to believe that Barletta was a Democrat and was more liberal. 
62\% of respondents guessed that he was a Democrat when he was posed next to African-Americans compared with 36\% to 44\% across all other categories of photos. 
Posing next to African-Americans also led non-white respondents to be more likely to agree that Barletta shared their values as can be seen in Figure~\ref{fig:sharevalues}.

\begin{figure}[!ht]
	\centering
	\includegraphics[width=.7\textwidth]{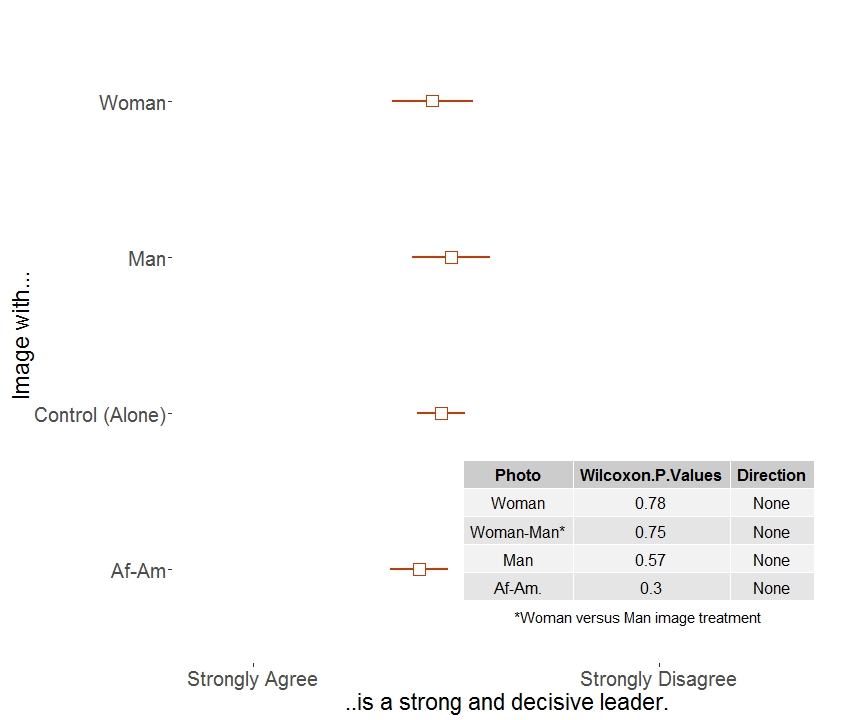}
	\caption{``Does the politician seem like a strong and decisive leader?'' Average agreement among respondents with 95\% confidence intervals.}
	\label{fig:strongleader}
\end{figure}

\begin{figure}[!ht]
	\centering
	\includegraphics[width=.7\textwidth]{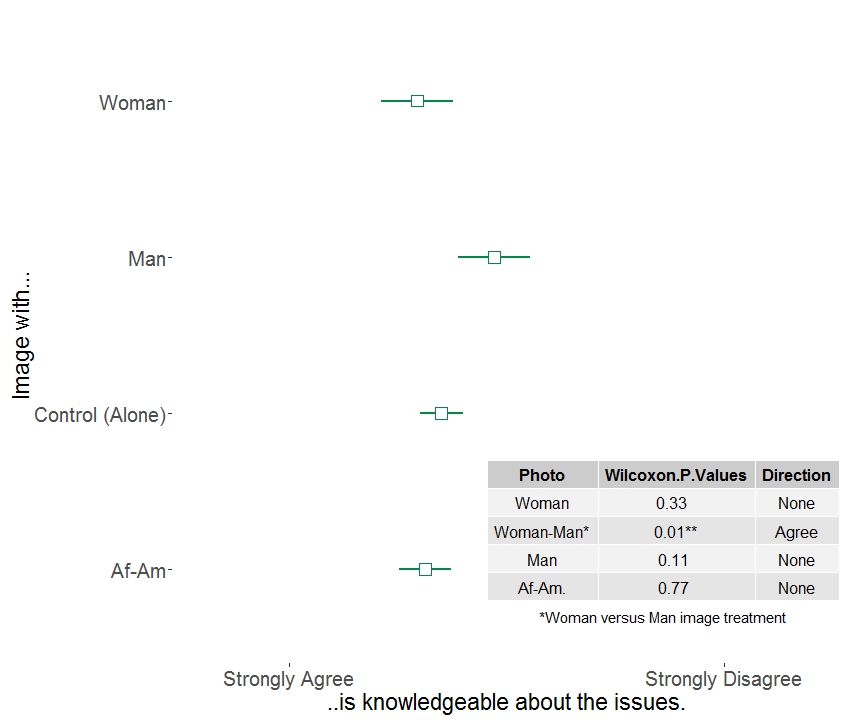}
	\caption{``Does the politician seem knowledgeable about the issues?'' Average agreement among respondents with 95\% confidence intervals.}
	\label{fig:knowledgable}
\end{figure}

Interestingly, we also find effects of gender on perceptions of qualification and competence. 
When Barletta poses next to a woman, respondents are more likely to agree that he is more trustworthy (Figure~\ref{fig:trustworthy}) and knowledgeable (Figure~\ref{fig:knowledgable}) than when he is alone or posing next to a white man.

\cleardoublepage

\section{Exploring photographic home styles with convolutional neural networks} \label{cnns}

The experiment discussed above demonstrates that the people whom a politician poses with affects how they are perceived on a number of dimensions related to home style.  
Here, we use a deep learning technique known as convolutional neural networks (CNNs) to conduct an empirical analysis of photographic home styles of members of the House and Senate.
Specifically, we are interested in exploring the following hypotheses related to racial representation and photographic home styles:

\begin{itemize}
	\item \textbf{H1:} Because Republican members of the House of Representatives do not rely on votes from minority constituents, they will not make efforts to signal identification and empathy with minority groups by posing with them in photos and/or posting these photos on social media.
	\item \textbf{H2:} Conversely, because Democrats tend to rely heavily on votes from minority groups~\citep{barreto2004mobilizing,cameron1996majority,denardo1980turnout}, they will make efforts to signal identification and empathy with minority groups by posing with them in photos and/or posting them on social media. 
	\item \textbf{H3:} As a consequence of \textbf{H1} and \textbf{H2}, we expect to find no relationship between Congressional district or state demographics and the racial composition of Facebook photos among Republicans in Congress but should find a strong relationship among Democrats.
 \end{itemize}

To test these hypotheses, we collected approximately 192,000 Facebook profile photographs from 230 members of the 114th House of Representatives and 52 members of the Senate\footnote{ See Appendix B for details about photo acquisition and analysis. For each member of the House and Senate, we acquired the entire set of photos from their Facebook profiles.}. 
We then build software which jointly leverages a large labeled yearbook photo database and photos from our Facebook database to build a 16-layer convolutional neural network which allows us to estimate the race of individuals within MCs Facebook photos. 
Finally, we calculate the racial demographics within MCs Facebook profiles and compare them to district demographics from American Community Survey 5-year estimates. 
Our data collection process and some background on convolutional neural networks is described in Appendix B and A, respectively, and  and the method that we used to build our software and the results are described below.

\subsection{Methods} 

\begin{figure}[!ht]
	\centering
	\includegraphics[width=\textwidth]{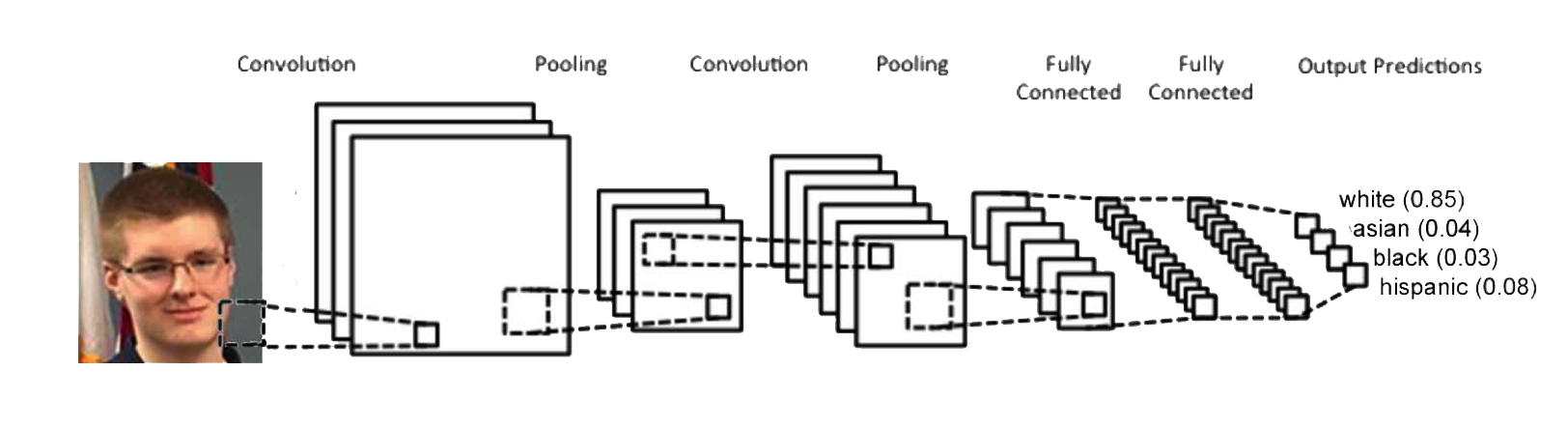}
	\caption{Depiction of a deep learning race classifier. The classifier utilizes pixel intensity data to estimate a complex non-linear function which can accurately predict high level features in images such as race.}
	\label{fig:cnn}
\end{figure}

We trained a race classifier for Facebook images based on ``VGG''~\citep{vgg} (depicted in Figure~\ref{fig:cnn} below), a deep classification network that was pre-trained on the ImageNet Large Scale Visual Recognition (ISLVRC) dataset~\citep{krizhevsky2012imagenet}. 
ISLVRC is a database containing a standard set of images used for bench-marking deep learning image classification models.
We fine-tuned the 16-layer VGG model on 17,500 PubFig~\citep{pubfig} training images using the race annotations provided with the dataset as ground truth.
Additionally, we included in the training set 44,000 portraits of American high school seniors from a large Yearbook dataset~\citep{ginosar2015century} that were classified with high confidence.

\begin{figure}[!ht]
	\centering
	\includegraphics[width=\textwidth]{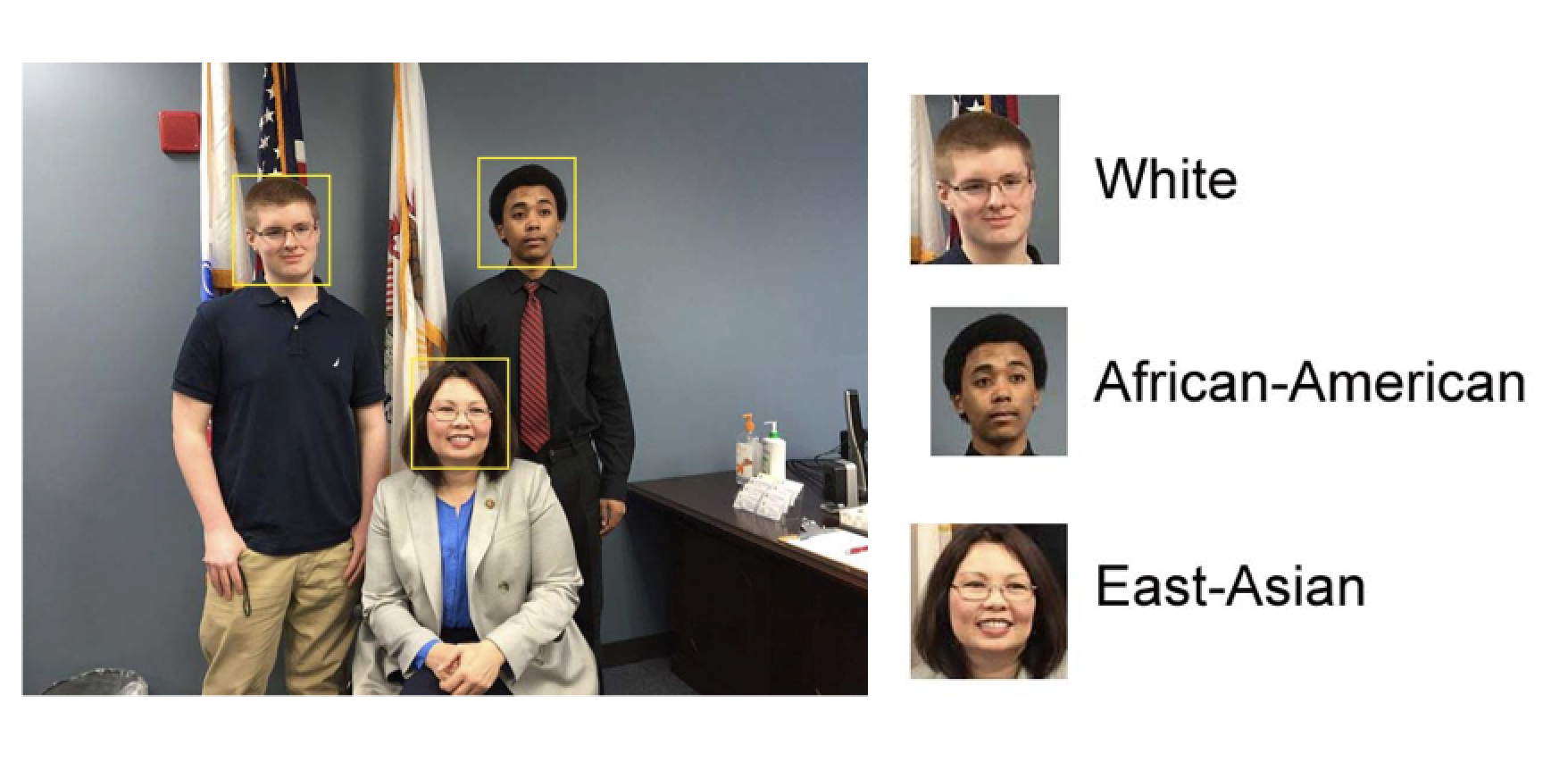}
	\caption{Faces in yearbook and Facebook images detected using Viola-Jones algorithm with Haarcascades, cropped and converted to grayscale as a pre-processing step.}
	\label{fig:cropped}
\end{figure}

All training images were cropped in a manner similar to Figure~ref{fig:cropped} by first detecting faces using Haarcascades~\citep{gorbenko2012face,lienhart2002extended}, a modified version of the Viola-Jones algorithm~\citep{viola2001rapid} to include only the face to match the Facebook test images.
Additionally all training and Facebook test images were converted to grayscale.
We split the training set into 61,000 training and 17,000 validation images, and used these to fine-tune the classifier for the labels White, African-American, Asian and Hispanic over 100,000 iterations.
Since training the classifier on the PubFig and Yearbook datasets and testing on Facebook portraits constitutes a domain shift that hurt the classification performance, we improved the classification accuracy via a process of bootstrapping by manually verifying the high-confidence race classifications and adding these into the training set.
Afterward, we further fine-tuned our classifier by training 20,000 iterations on the augmented training set.
The final average cross-validated accuracies were 90\% for Whites, 85\% for African-American, 75\% for Asian and 65\% for Hispanic.

\subsection{Results}

\begin{figure}[!ht]
	\centering
	\includegraphics[width=.8\textwidth]{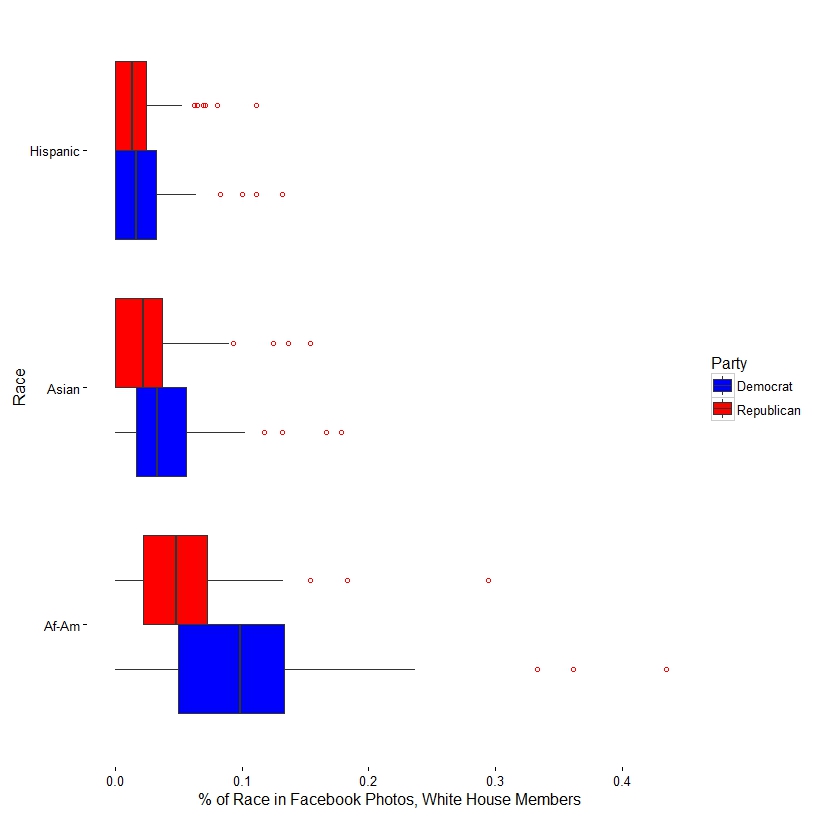}
	\caption{Proportion African-American, Hispanic and Asian in Facebook photos of white House of Representative Members.}
	\label{fig:pctminority-house}
\end{figure}

After classifying the race of individuals in House and Senate members' Facebook profiles, we test the hypotheses discussed above by estimating Facebook profile ``demographics'' for each member of Congress and compare these demographics to district and state demographics from American Community Survey (ACS) 5-year estimates.
 We begin our analysis by exploring the raw estimated proportion of African-Americans, Hispanic and Asians in the Facebook profiles of white Republican and Democratic House and Senate members.

\begin{figure}[!ht]
	\centering
	\includegraphics[width=.8\textwidth]{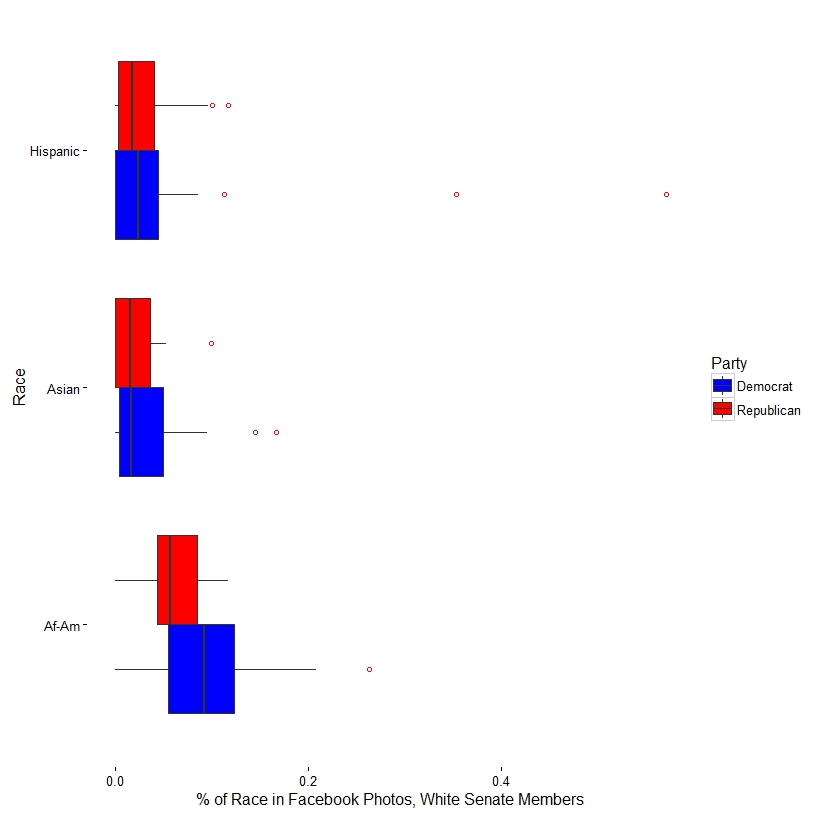}
	\caption{Proportion African-American, Hispanic and Asian in Facebook photos of white Senate Members.}
	\label{fig:pctminority-senate}
\end{figure}

Figures~\ref{fig:pctminority-house} and~\ref{fig:pctminority-senate} are box plots of the raw estimated proportion African-American, Hispanic and Asians in the Facebook profiles of white representatives in the House and Senate, respectively.
The clear and consistent pattern that emerges from these figures is that white Democrats in both chambers appear to include a significantly higher proportion of African-Americans and Asians in their images, but similar proportions of Hispanic voters.
This provides some evidence for our hypothesis that race is an essential component of Democrat's, but not Republican's home styles, especially since white representatives in both chambers exhibit roughly the same patterns.

The question that remains, however, is whether these differences are due to chance or whether they involve strategic behavior.
Indeed, Democratic reps have higher minority proportions in their districts and states and may just randomly happen to take more photos with minorities as they encounter them at the various events that they attend.
If this were true, we should observe clear and roughly equivalent correlations between district/state demographics and Facebook racial demographics among Democrats and Republicans.
Evidence of strategic behavior, on the other hand, would manifest as differences among Democrats and Republicans in terms of the relationship between district/state demographics and Facebook racial demographics.

Specifically, we would expect to find that Democrats will use the race of individuals in the photos that they post as a means of inspiring trust among  minority constituents. 
If the goal of Democratic MCs is to gain trust through engendering identification and empathy with minority constituents, this would be reflected in the photographic home styles of Democratic MCs as a strong and positive relationship between Facebook photo racial demographics and district demographics.
Republicans, on the other hand, have little electoral incentives to appeal to minority groups through photos and as a result are unlikely to use race as a means of inspiring trust among constituents.
Thus, we would expect a weak/non-existent relationship between Facebook racial demographics and district racial demographics among Republican MCs. 
We find strong evidence of strategic behavior in the use of photos among Democrats and Republicans as we demonstrate below with analyses comparing Facebook photo racial demographics with actual 2015 American Community Survey (ACS) 5-year estimates of Congressional district and state Census demographics.
 
\begin{figure}[!ht]
	\centering
	\includegraphics[width=\textwidth]{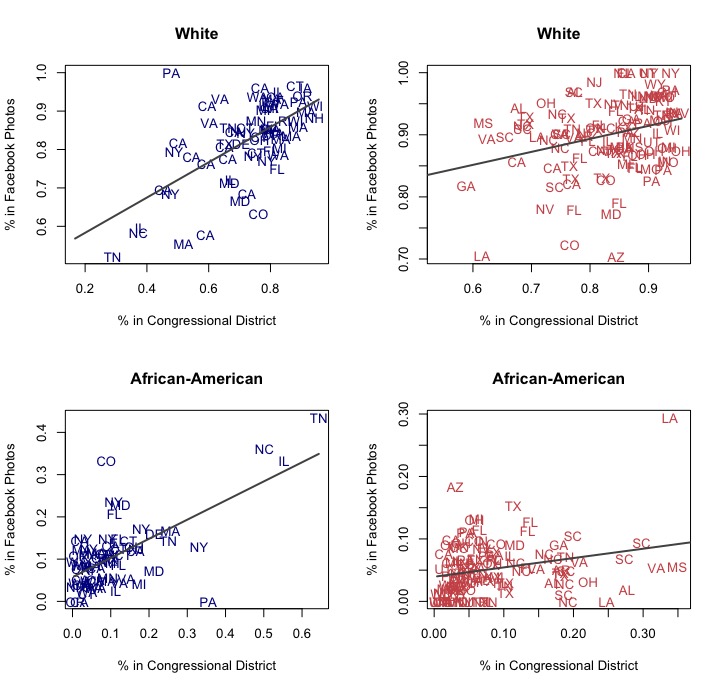}
	\caption{House results: \% white and \% African-American in House district v. Facebook photos. Democrats $=$ \textit{blue}, Republicans $=$ \textit{red} (white House members).}
	\label{fig:districtfb}
\end{figure}

Figure~\ref{fig:districtfb} contains plots of Congressional district demographics from 2015 ACS 5-year estimates versus Facebook profile demographics produced by our convolutional neural network classifier among white House members only. 
There is clearly a strong relationship between district racial demographics and Facebook photo demographics among Democratic House members but not Republicans, suggesting that House members photographic home styles are shaped by electoral pressures.
These results are also consistent across racial categories when state fixed effects are included as shown in Figure~\ref{fig:districtfbfe} in Appendix D.

\begin{figure}[!ht]
	\centering
	\includegraphics[width=\textwidth]{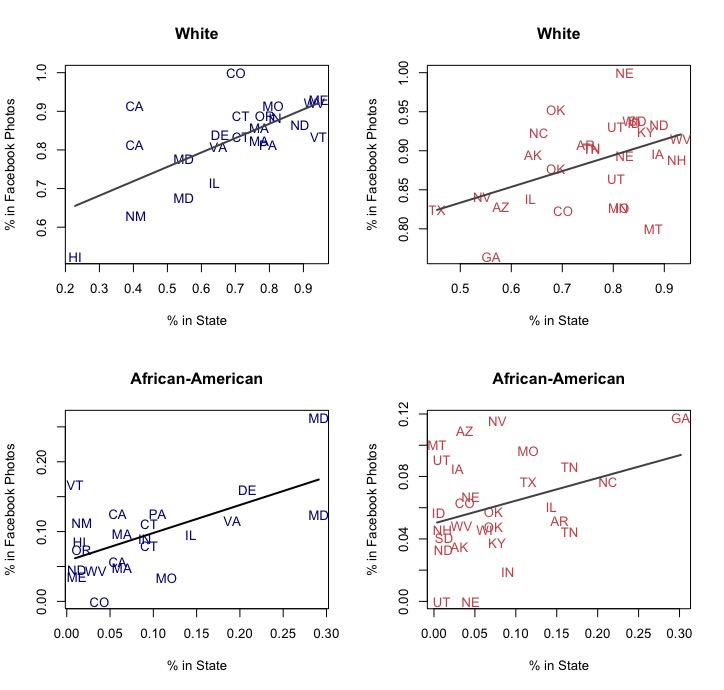}
	\caption{Senate results: \% white and \% African-American in State v. Facebook photos. Democrats $=$ \textit{blue}, Republicans $=$ \textit{red} (white Senators).}
	\label{fig:statefb}
\end{figure}

Figure~\ref{fig:statefb} contains plots of 2015 Census State demographics  versus Facebook profile demographics produced by our convolutional neural network classifier among white Senators.
The patterns are similar the House results from Figure~\ref{fig:districtfb} with a slightly stronger relationship among Republicans.

\cleardoublepage

\section{Discussion} \label{discussion}

The use of images by political figures to manipulate public opinion and sentiment is by no means a new phenomenon.
There is no question that Hitler's rise to power was facilitated by director and Nazi propagandist Leni Riefenstahl's films about him while some of the same directorial techniques were used by U.S. and Allied forces to encourage people to buy bonds and join the Army as part of the war effort.
As discussed above, shrewd political figures such as Lyndon Johnson recognized the potential that images had to shape how they were perceived by the public and accordingly appointed the first White House photographer to do just that during his term in office.
At the same time, however, photographs taken by brave journalists and chroniclers of the Vietnam War laid bare the horrors and devastation of a war which eventually led to Johnson's steep decline in popularity and his eventual decision to not run for a second term.

While the use of images to achieve political ends is not new, our ability to systematically study and understand how, when and why they are used has only recently been made possible by the prolific use of images by politicians and members of their staff made possible by the internet and recent developments in computer vision which allow for quick identification of complex features from labeled image data using only the pixels themselves.
Here, we take advantage of both developments in an effort to provide a broad framework for political image analysis using these techniques and simultaneously demonstrate how they can be used to understand the modern relevance of home style~\citep{fennohomestyle}.
By breaking down images into their simplest political elements: objects, people and poses, our framework provides a basis from which scholars can explore which aspects of images are used by politicians and others for the purpose of influencing opinion. 

In addition to this, we make significant contributions to the understanding of home style in the 21st century.
We first develop a theory of photographic home style by linking our broader image analysis framework to Fenno's main conceptual elements of home style.
We then demonstrate that the characteristics of individuals that politicians pose with affect how they are perceived across a number of politically relevant dimensions related to home style which include qualification, identification and empathy.
Using deep learning software that we developed with convolutional neural networks and an empirical analysis of photos posted by members of the House and Senate on their Facebook profiles, we then demonstrate that electoral pressures shape photographic home style. 
Specifically, we discover that Democratic members of Congress, either intentionally or unintentionally, match the racial demographics of their Facebook profiles to their district demographics in what we hypothesize is an attempt to engender identification and empathy among minority constituents in their districts.
Republican members of Congress, however, do not employ this strategy with respect to race but may employ a similar strategy using members of constituency groups, such as veterans, that are more likely to contribute to their reelection success. 
We hope that future research on this topic will explore these and other possibilities.

\doublespacing
\newpage
\bibliographystyle{apsr_fs} 
\bibliography{citations}

\newpage

\section*{Appendix A: Computer Vision Background}

\subsection{An Overview of Computer Vision} 

Computer vision is an area of research in computer science whose primary task is concerned with understanding how images are represented on machines and using this information to develop algorithms which can identify ``high-level'' features in images such as people, places, objects, poses and symbols~\citep{girshick2014rich, klette2014concise, le2013building}.
From the perspective of social science research, it involves the identification of semantic properties of images as represented by image features which modern computer vision techniques make possible~\citep{joo2014visual}.
Until recently, however, the identification of high level features within images was practically impossible. 
This was largely due to a focus on mathematical rigor and image geometry in the earlier years of computer vision research which was in turn the result of computational and hardware limitations faced by researchers in this field~\citep{forsyth2002computer}. 

\subsubsection{Image as Data}

\begin{figure}[!h]
	\centering
	\includegraphics[width = \textwidth]{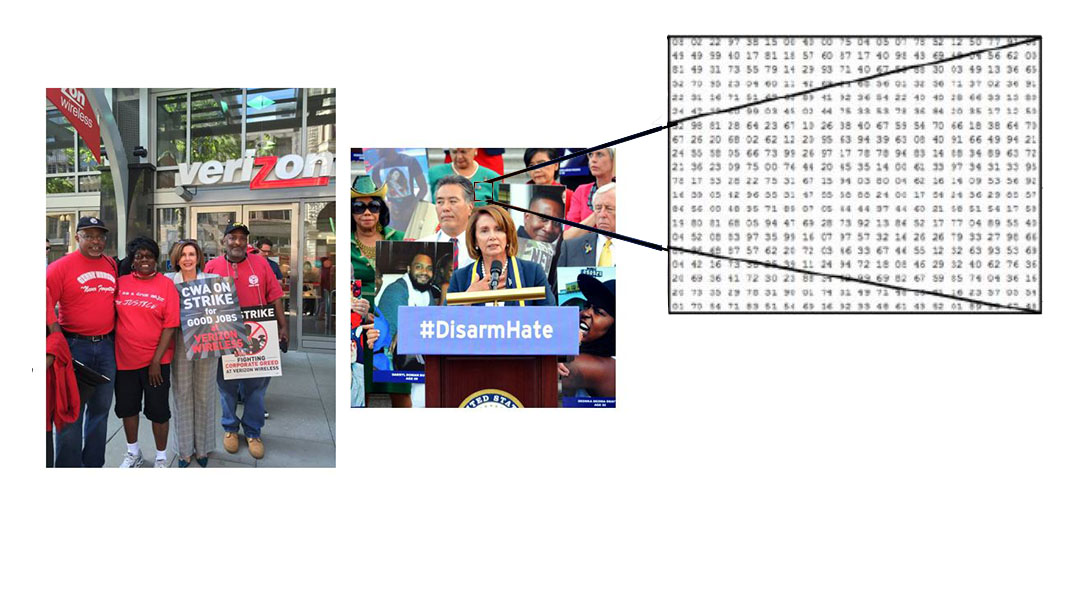}
	\caption{Images of Nancy Pelosi as represented on a machine by a matrix of pixel intensities.}
	\label{fig:dtrump2}
\end{figure}

When we view an image on a computer screen, we are seeing a collection of pixels stacked in a certain order. 
Grayscale (black and white) images are represented on a machine as a single matrix of pixel intensity values which represent the brightness of the image. 
The most common format for pixel intensity values is the \textit{byte image} which is stored as an 8-bit integer that takes on a range of integer values between 0 and 255. 
Color images are also stored as pixel intensity values, but instead of containing pixels values across a light/dark dimension, color images contain pixel intensity values across red, green and blue channels. 
Pixel intensity values for both grayscale and color images are thus represented as a one or three matrices of pixel intensity values in the range of 0 to 255.
Because images, both color and grayscale, are represented on machines as either one or three matrices, each image in a machine is typically represented as a series of \textit{tensors} or arrays of multidimensional arrays. 

\begin{figure}[!ht]
	\centering
	\includegraphics[width =\textwidth]{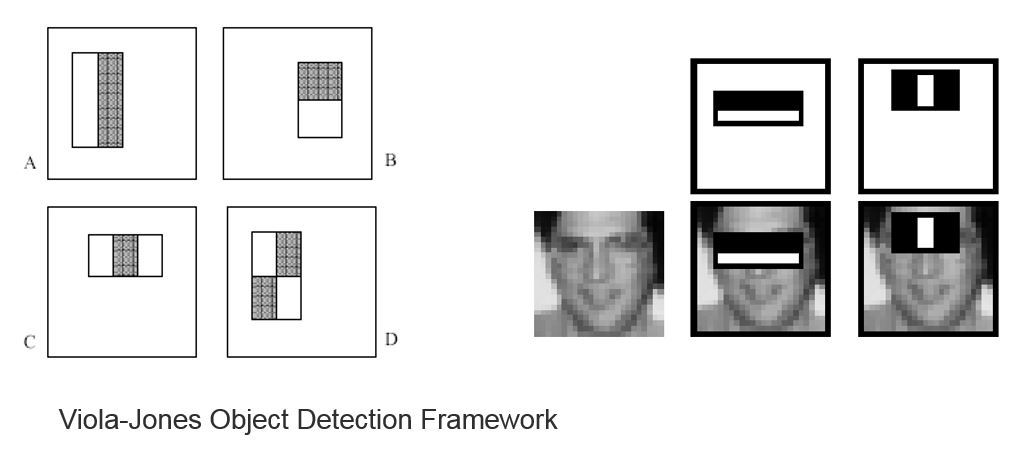}
	\caption{The Viola-Jones algorithm is a successful example of  an older computer vision algorithm which used mathematical models and object geometry to identify image features.}
	\label{fig:violajones}
\end{figure}

Figure~\ref{fig:dtrump2} provides a more concrete example of how an image is represented on a machine. 
This picture was originally a 640x412 color pixel image. 
On a computer, it is represented as three 640x412 arrays or a 791,040x1 vector.
This example makes it clear why images are fundamentally ``big data'' when represented as data on a computer. 
Even the small Nancy Pelosi images in Figure~\ref{fig:dtrump2} are represented by a comparatively large amount of numerical data. 
As the size and number of images to be analyzed increases, it’s easy to see how the hardware and software requirements to perform even simple operations such as matrix inversion via Cholesky decomposition algorithms which tend to be $O(N^3)$~\citep{bertsekas1999nonlinear}, would increase dramatically even for analyzing a small image dataset.  
Because of this ``big data'' nature of images and hardware limitations in the past, early pioneers in the field of computer vision focused on conceptualizing images as sets of Platonic solids and used mathematical models to identify high-level features in images such as faces and objects.

\begin{figure}[!h]
	\centering
	\includegraphics[width =.8\textwidth]{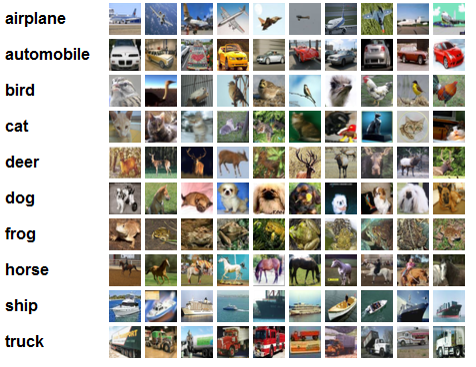}
	\caption{CIFAR-10 are a dataset of benchmarks for image classifiers containing a wide variety of images within 10 classes. \textit{Source: \href{https://www.cs.toronto.edu/~kriz/cifar.html}{https://www.cs.toronto.edu/~kriz/cifar.html}}}
	\label{fig:cifar10}
\end{figure}

Unfortunately, as the result of natural variation common to images such as lighting, camera angles and occlusion, not to mention natural variation in \textit{types} of simple geometric objects such as chairs and tables, most efforts to identify features of images using mathematical modeling, with the exception of faces, were largely unsuccessful. 
Although this original focus led to robust face detection methods such as the Viola-Jones algorithm which is shown in Figure~\ref{fig:violajones}, these algorithms largely failed to detect other types of basic objects such as automobiles, airplanes, birds, cats etc. which currently form the basis of image benchmarks such as the CIFAR-10. 
It was not until larger annotated datasets of images were available and computational power improved significantly that research employing data-heavy approaches such as artificial neural networks were able to perform spectacularly on a number of benchmarks spanning many categories. 
Specifically, an artificial neural network which employs pixel data convolutions that has come to be known as a convolutional neural network, the method utilized in this paper, has opened the door to image analysis in social science~\citep{lecun2015deep}.		

\subsection{Overview of Convolutional Neural Networks}

Convolutional neural networks (CNNs), as mentioned above, are a type of artificial neural network employed in image analysis with high rates of success for image classification tasks.
Like all artificial neural networks, they are comprised of ``layers'' which are just simple functions estimated from data that are then ``pieced together'' to form more complex non-linear functions of the kind required to perform complicated tasks such as identifying features from pixel intensity data alone.
In functional form, a neural network is simply the composition of many functions $f()$ with associated weights \textbf{w}, layers \textbf{L}, and data $\mathbf{\theta}$.

\begin{equation}\label{eq:nnfunction}
	f(\mathbf{\theta}) = f_{L}(...f_{2}(f_{1}(\mathbf{\theta}, \mathbf{w}_{1}), \mathbf{w}_{2})...), \mathbf{w}_{L}) 
\end{equation}

Weights for $f(\mathbf{\theta})$ are estimated such that they minimize the difference between predicted and actual class labels as provided by the data by minimizing the loss function $\lambda(\mathbf{z_{i}},f(\mathbf{\theta_{i}},\mathbf{w}))$ where $\mathbf{z_{i}}$ are the true, labeled values and $f(\mathbf{\theta_{i}},\mathbf{w}))$ are the predicted values as produced by the model. 
The empirical loss is then  the average loss over each of the examples $i$:

\begin{equation}\label{eq:emploss}
\displaystyle	\Lambda(\mathbf{w}) = \frac{1}{n} \sum_{i = 1}^{n} \lambda(\mathbf{z_{i}},f(\mathbf{\theta_{i},\mathbf{w})})
\end{equation}

The loss function is minimized through an optimization technique known as \textit{back-propagation} which continuously updates the loss function through a series of \textit{forward} and \textit{backward} passes through the model. 
During the forward pass, the data is first passed through the model in Equation~\ref{eq:nnfunction} with a randomly initialized set of weights and predicted values are generated.
Gradients for the weights in each of the layers are then computed and adjusted accordingly:

\begin{equation}\label{eq:backprop}
	\frac{\partial f}{\partial \mathbf{w}_{l}} = \frac{\partial}{\partial \mathbf{w}_{l}} f(\mathbf{\theta}) 
\end{equation}

What differentiates \textit{convolutional neural networks} from artificial neural networks in general is that all of the functions from equation~\ref{eq:nnfunction} are operators that are translation invariant. 
In words, the convolutional aspect of convolutional neural networks \textit{slide} the set of functions across portions of pixel data in the image similar to the way in which basic filters, such as a Gaussian blur, work to transform images. 

\section*{Appendix B: Image Acquisition and Analysis}

Images from each member of Congress were acquired through the following means. The authors acquired the Facebook usernames of each member of Congress from a Github database of members of Congress social media accounts which can be found in a $.yaml$ formatted file found here: \href{https://github.com/unitedstates/congress-legislators/blob/master/legislators-social-media.yaml}{https://github.com/unitedstates/congress-legislators/blob/master/legislators-social-media.yaml}.
In order to gain access to photos from each members profile, the authors registered with Facebook's API and used Python code to iteratively download photos from each members Facebook profile.
This resulted in 192,000 files.

In order to facilitate analysis of these photos, a facial recognition algorithm was run through the the photo database and only photos with faces were identified. 
A 10\% sample of these photos from each member of Congress was then randomly sampled for analysis yielding the final classified image database discussed above.

\section*{Appendix C: Image Experiment Recruitment and Analysis}

\subsection{Experimental Design}

The survey experiment discussed in Section~\ref{experiment} was designed using Qualtrics survey technology. 
Each of the 7 image treatments were randomized using Qualtrics proprietary randomizer.

\subsection{Respondents}

Respondents were recruited through Amazon's Mechanical Turk via a HIT advertised as a ``3-5 minute survey of your opinions on a policy question.''
Respondents were restricted to those in the United States with HIT approval rates greater than or equal to 99\% with the number of HITs approved greater than or equal to 1,000. 
Respondents were provided $ \$0.50$ compensation for taking the survey. 
Survey data was collected once on July 08, 2016 and again on October 19, 2016.
A total of 1,150 responses to the survey were collected.
78 respondents failed basic validation questions and were excluded from the study yielding a total of 1,072 respondents used for the analyses.

Breakdowns of respondent demographics with respect to party identification, age, gender, race , education and employment status are as follows.
	
\begin{table}
	\centering \footnotesize
	\begin{tabular}{lll}
					  & \textbf{Mean/\%} & \textbf{95\% CI} \\ \hline
		 \textbf{Age} & 36.54 	& (35.86, 37.22)  \\
		 \textbf{Male} & 56.6\%  &  (53.51\%, 59.48\%)  \\
		 \textbf{White} & 78.04\% & (75.54\%, 80.53\%)   \\
		 \textbf{African-American} & 5.75\% & (4.34\%, 7.15\%) \\
		 \textbf{Hispanic} & 6.13\% & (4.68\% ,7.57\%)   \\
		 \textbf{Some College or College Graduate}  & 88.84\% & (86.93\%,90.73\%) \\ 
		 \textbf{Unemployed} & 6.97\% & (5.43\%,8.50\%) \\ 
		  \textbf{Democrat} & 42.28\% & (39.30\%,42.25\%) \\
		  \textbf{Republican} & 18.36\% & (16.03\%, 20.69\%) \\
		  \textbf{Independent} & 34.93\% & (32.06\%,37.80\%) \\	\hline\hline
	\end{tabular}
	\caption{Respondent demographics from the survey experiment described in Section~\ref{experiment}}
\end{table}

As is typical of many Mechanical Turk samples, respondents are in their 30's,slightly more male, white and identify with the Democratic party or as independents. 

\cleardoublepage

\section*{Appendix D: Fixed Effects Models}

\begin{figure}[!ht]
	\centering
	\includegraphics[width=.8\textwidth]{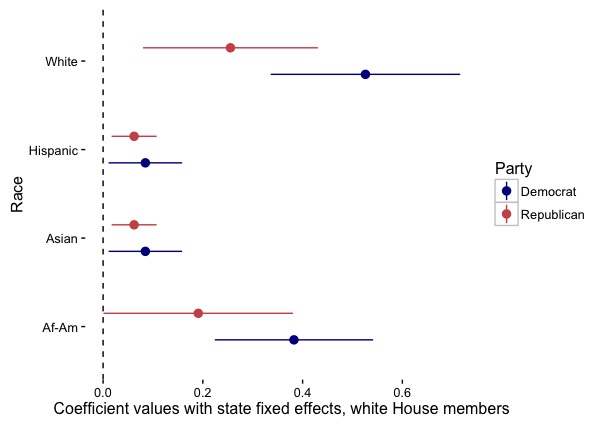}
	\caption{House results: plot of coefficient values for state fixed effects regressions where $Y =$ \% (White, Black, Hispanic, Asian) in district and $X =$ \% (White, Black, Hispanic, Asian) in Facebook photos. Higher coefficient values indicate that district demographics better predict Facebook photo demographics.}
	\label{fig:districtfbfe}
\end{figure}

\end{document}